\documentclass[runningheads]{llncs}

\usepackage[utf8]{inputenc}
\usepackage{algorithm}
\usepackage[noend]{algpseudocode}
\usepackage{amssymb}
\usepackage{cite}
\usepackage{paralist}
\usepackage[portuguese]{babel}
\usepackage{graphicx}
\usepackage{subfig}

\algblockdefx{Payload}{EndPayload}[1]{\textbf{payload} #1}{end}
\algblockdefx{Query}{EndQuery}[4]{\textbf{query} \emph{#1} (#2) : #3 \emph{#4}}{end}
\algblockdefx{Update}{EndUpdate}[2]{\textbf{update} \emph{#1} (#2)}{end}
\algblockdefx{UpdateReturn}{EndUpdateReturn}[4]{\textbf{update} \emph{#1} (#2) : #3 \emph{#4}}{end}
\algblockdefx{Merge}{EndMerge}[3]{\textbf{merge} (\emph{#1}, \emph{#2}) : payload \emph{#3}}{end}
\algblockdefx{Compare}{EndCompare}[3]{\textbf{compare} (\emph{#1}, \emph{#2}) : boolean \emph{#3}}{end}
\algblockdefx{AtSource}{EndAtSource}[1]{\textbf{atSource} (#1)}{end}
\algblockdefx{AtSourceReturn}{EndAtSourceReturn}[3]{\textbf{atSource} (#1) : #2 \emph{#3}}{end}
\algblockdefx{AtSourceMultipleReturn}{EndAtSourceMultipleReturn}[2]{\textbf{atSource} (#1) : #2}{end}
\algblockdefx{Downstream}{EndDownstream}[1]{\textbf{downstream} (#1)}{end}

\newcommand{\Pre}{\textbf{pre} }
\newcommand{\Let}{\textbf{let} }
\algnewcommand{\IfThenElse}[3]{\State \algorithmicif\ #1\ \algorithmicthen\ #2\ \algorithmicelse\ #3}

\algnotext{EndPayload}
\algnotext{EndQuery}
\algnotext{EndUpdate}
\algnotext{EndUpdateReturn}
\algnotext{EndMerge}
\algnotext{EndCompare}
\algnotext{EndAtSource}
\algnotext{EndAtSourceReturn}
\algnotext{EndAtSourceMultipleReturn}
\algnotext{EndDownstream}

\newcommand{\addwins}{\emph{add-wins}}
\newcommand{\remwins}{\emph{remove-wins}}
\newcommand{\add}{\emph{add}}
\newcommand{\rem}{\emph{remove}}
\newcommand{\addop}{\emph{add}}
\newcommand{\remop}{\emph{remove}}
\newcommand{\remwinsop}{\emph{removeWins}}
\newcommand{\addope}{\emph{add}($e$)}
\newcommand{\remope}{\emph{remove}($e$)}
\newcommand{\remwinsope}{\emph{removeWins}($e$)}

\makeatletter
\renewcommand{\ALG@name}{Algoritmo}%
\renewcommand{\keywordname}{{\bf Palavras-chave:}}
\makeatother

\begin{document}
\title{\emph{Set CRDT} com Múltiplas Políticas de Resolução de Conflitos}

\author{André Rijo \and
Carla Ferreira \and
Nuno Preguiça}


\institute{NOVA LINCS \& DI, FCT, Universidade NOVA de Lisboa}
    
\maketitle


\begin{abstract}
Um CRDT é um tipo de dados que pode ser replicado e modificado concorrentemente sem coordenação,
garantindo-se a convergência das réplicas através da resolução automática de conflitos.
Cada CRDT implementa uma política específica para resolver conflitos. 
Por exemplo, um conjunto CRDT \emph{add-wins} dá prioridade ao \add{} 
aquando da execução concorrente dum \add{} e \rem{} do mesmo elemento.
Em algumas aplicações pode ser necess\'ario usar diferentes políticas para 
diferentes execuções de uma operação -- por exemplo, uma aplicação 
que utilize um conjunto CRDT \addwins{} pode querer que alguns \emph{removes} ganhem sobre \emph{adds} concorrentes.
Neste artigo é apresentado e avaliado o desenho dum conjunto CRDT que implementa as semânticas referidas.

\end{abstract}


\section{Introdução}

\vspace{-2mm}
Hoje em dia muitos serviços de larga escala são hospedados na Internet, tendo alguns deles milhões de utilizadores diários.
De modo a tolerar falhas e providenciar acesso rápido, esses serviços precisam de replicar os seus dados por diversos centros de dados.
Num mundo perfeito, os serviços teriam consistência forte, alta disponibilidade e tolerância a partições da rede.
Infelizmente, o 
teorema CAP \cite{gilbert2002brewer} afirma que é impossível fornecer estas três garantias simultaneamente.

Tendo em conta esta limitação, muitos sistemas preferem focar-se em fornecer alta disponibilidade e tolerar partições na rede, sacrificando a consistência forte.
Tipicamente esses sistemas oferecem antes uma forma de consistência fraca denominada de consistência eventual, a qual apenas garante que, eventualmente, todas as réplicas convergem para o mesmo estado.
Este modelo permite que as operações sejam executadas concorrentemente em diferentes réplicas, sem ser necessário sincronização. Contudo, as operações concorrentes precisam de ser combinadas de modo a garantir convergência de estado.

Uma possível solução é usar os Tipos de Dados Replicados sem Conflitos (CRDTs) \cite{shapiro2011comprehensive}, sendo que estes fornecem por construção uma política de resolução de conflitos. Os CRDTs permitem que as operações executem localmente na réplica fonte e que estas sejam propagadas assincronamente para as réplicas remotas.
Existem vários CRDTs que representam tipos de dados diferentes, como por exemplo contadores, mapas, conjuntos, listas 
\cite{shapiro2011comprehensive, preguica2009commutative}.

Os CRDTs fornecem uma política de resolução de conflitos determinística de modo a poderem lidar com operações concorrentes que não comutam de forma natural.
Por exemplo, um conjunto CRDT providência duas operações: \addope{} adiciona o elemento $e$ ao conjunto; e \remope{} remove o elemento $e$ do conjunto.
Contudo, as operações de adição e remoção do mesmo elemento não são comutativas.
Como tal, é necessário arbitrar um estado final para o objeto para quando essas operações são executadas concorrentemente.
Uma possível semântica de concorrência, conhecida como \addwins{} (resp. \remwins{}), é de dar prioridade à operação de adição (resp. remoção), ou seja, o elemento ficar presente (resp. não ficar presente) no conjunto.


Apesar de uma política \addwins{} poder ser adequada para algumas aplicações outras podem, contudo, preferir uma política \remwins{}. 
Outras podem ainda querer a possibilidade de usar ambas as políticas para diferentes elementos ou situações.
Por isso, neste trabalho, propomos um conjunto CRDT que fornece, simultaneamente, as políticas \addwins{} e \remwins{}.

\vspace{-3.0mm}
\subsection{Exemplos de utiliza\c{c}\~ao}
\label{subsec: exemplo}
\begin{figure}
\vspace{-5mm}
\centering
\includegraphics[width=0.42\textwidth]{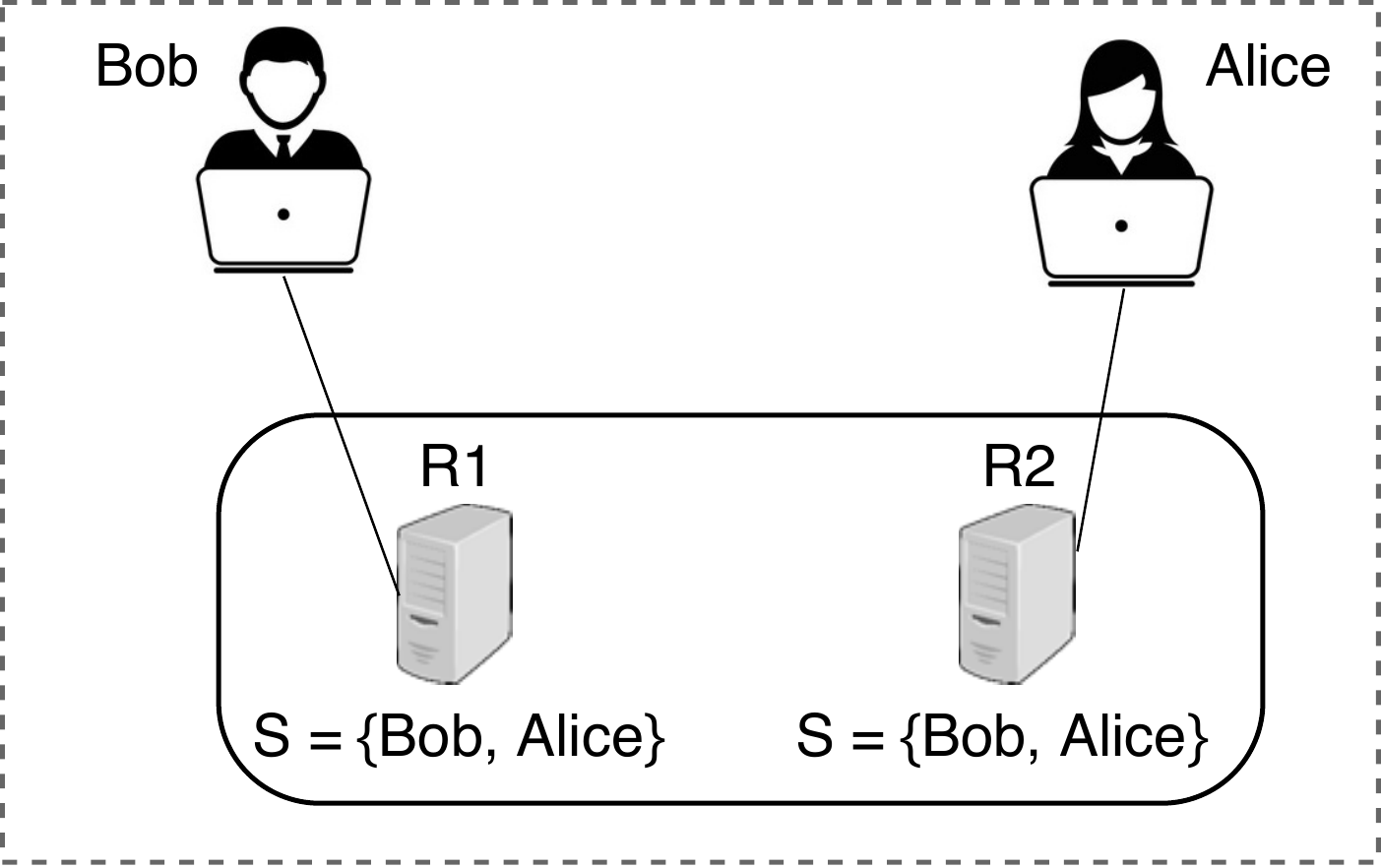}
\vspace{-1mm}
\caption{Serviço de conversa online, com dois utilizadores e dois servidores. O conjunto \emph{S} representa os utilizadores online.} 
\label{fig_chat_1}
\vspace{-5mm}
\end{figure}

\paragraph{Serviço de conversa online.}
Para ilustrar a aplicabilidade do  CRDT proposto, considere-se um serviço que disponibiliza uma sala de conversa online

O serviço mantém nas várias réplicas um conjunto \emph{S} que indica quais os utilizadores que estão atualmente online, sendo apenas necessário cada cliente estar ligado a um dos servidores para ser considerado como online.
Quando um utilizador se (re)conecta a um servidor, é realizado um \add{} em \emph{S}. Um \rem{} é executado quando a conexão é perdida ou o utilizador faz \emph{logout}.

\begin{figure}
\centering
\includegraphics[width=0.82\textwidth]{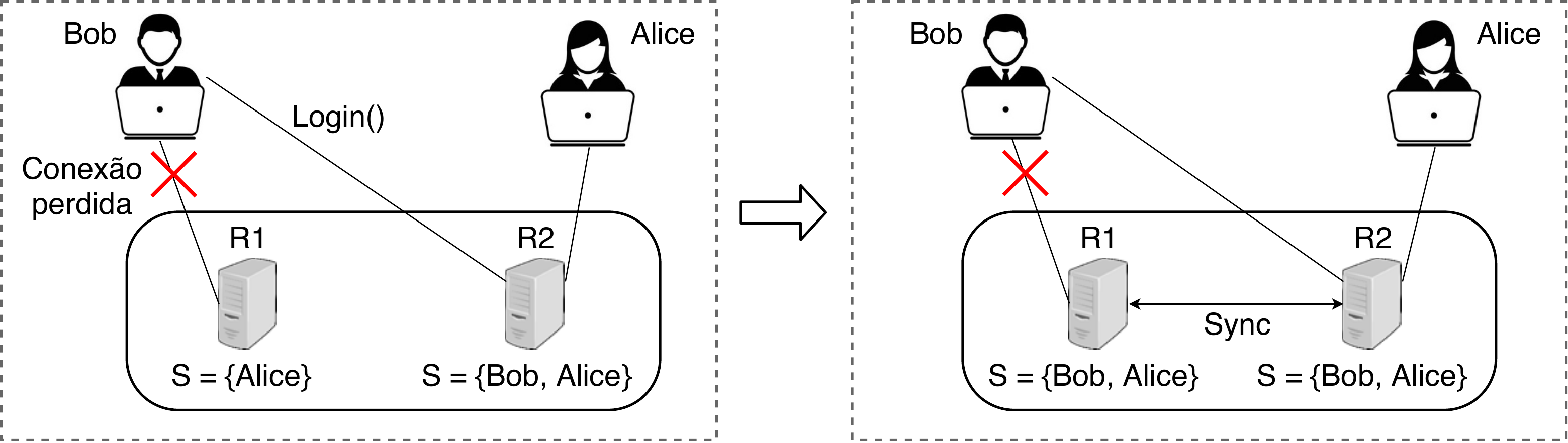}
\vspace{-1mm}
\caption{Situação em que existe uma perda de ligação e uma nova ligação concorrentemente. Aplicando a política \addwins{}, o Bob continua online.} \label{fig_chat_2}
\vspace{-1mm}
\end{figure}

\begin{figure}
\centering
\includegraphics[width=0.82\textwidth]{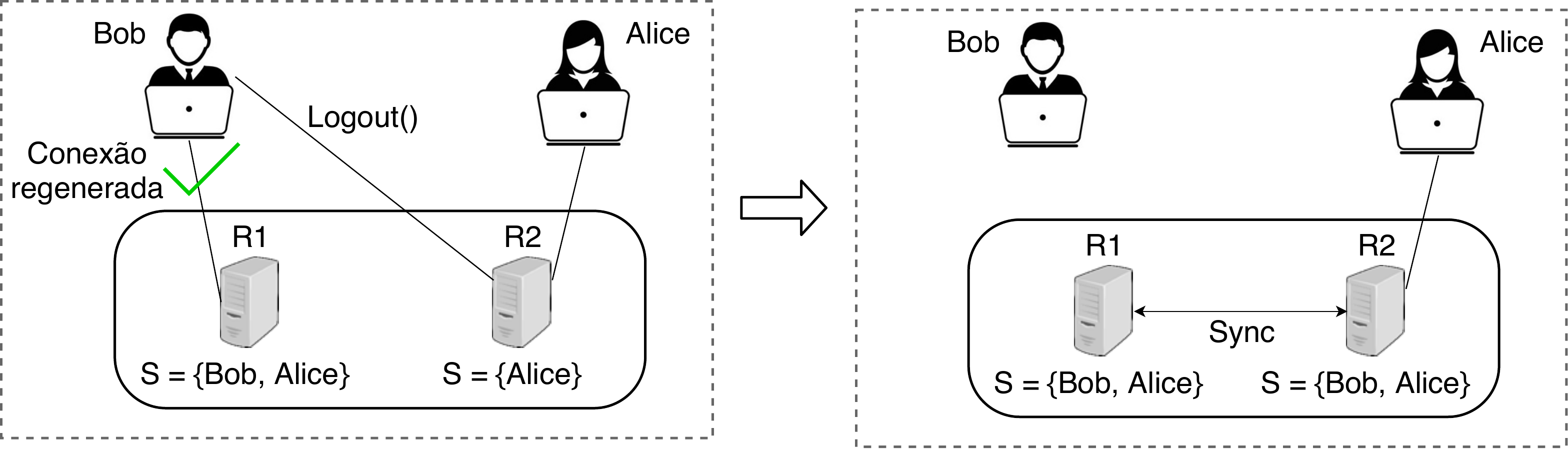}
\vspace{-1mm}
\caption{Situação em que existe uma regeneração de uma ligação e um \emph{logout} concorrentemente. Aplicando a política \addwins{}, o Bob continua online.} \label{fig_chat_3}
\vspace{-1mm}
\end{figure}

Considere-se agora as seguintes situações:
\vspace{-1.0mm}
\begin{itemize}
\setlength{\itemsep}{0.25em}
    \item A conexão do Bob à replica \emph{R1} é perdida (\rem{}). Concorrentemente, o Bob liga-se à replica \emph{R2} (\add{}). Quando as réplicas sincronizarem, se for aplicada a política \addwins{}, o Bob fica como online, conforme é pretendido (Figura \ref{fig_chat_2}). Se fosse aplicada a política \remwins{}, o Bob ficaria como offline, o que não seria correto;
    \item Depois da situação acima referida, a conexão do Bob à replica R1 regenera (\add{}). Concorrentemente, o Bob realiza um \emph{logout} na réplica R2 (\rem{}). Quando as réplicas sincronizarem, se for aplicada a política \addwins{} o Bob fica como online, mesmo já não estando a utilizar o serviço (Figura \ref{fig_chat_3}). Neste cenário, seria desejável que fosse aplicada a política \remwins{}.
\end{itemize}
Em suma, as duas situações constituem um caso onde é vantajoso ter um conjunto CRDT que forneça tanto a política \addwins{} como \remwins{}.

\vspace{-1.0mm}
\paragraph{Serviço com super utilizador.}
Outro cenário no qual um conjunto CRDT com ambas as políticas é útil consiste num serviço 
no qual existam utilizadores especiais cujas operações têm maior 
prioridade em relação às dos outros utilizadores. 
%
Para exemplificar este cenário, considere-se um serviço que controla quais os utilizadores 
que têm acesso a uma pasta partilhada pelo dono da mesma. 
Assumindo que se pretende que certos utilizadores (definidos pelo dono) possam convidar outros, 
é desejável que quando o dono remove o acesso a um utilizador, esta remoção não seja sobreposta por uma adição desse utilizador. 
Por outro lado, quando dois utilizadores que não sejam o dono adicionam e 
removem concorrentemente o mesmo utilizador, é preferível que esse utilizador fique com acesso à pasta.
Em suma, um conjunto CRDT com ambas as políticas é adequado a situações onde se  pretenda
que certos utilizadores (elementos) tenham prioridade diferente dos restantes utilizadores (elementos) aquando da execução das suas operações.

\vspace{-3mm}
\subsection{Contribuições}
Tendo como motivação o exemplo apresentado na Secção \ref{subsec: exemplo}, neste trabalho fazemos as seguintes contribuições:
\vspace{-1.0mm}
\begin{itemize}
\setlength{\itemsep}{0.25em}
	\item A proposta de um mecanismo para suportar múltiplas políticas no mesmo CRDT, através da adição de operações extra;
	\item Duas definições state-based (uma delas otimizada) e op-based de um conjunto CRDT que fornece as políticas \addwins{} e \remwins{};
	\item Uma avaliação experimental do CRDT proposto, sendo comparado o seu desempenho  com CRDTs similares.
\end{itemize}

O resto do artigo organiza-se nas seguintes secções: as Sec\c{c}\~oes \ref{sec: especificacao} e  \ref{sec:avaliacao} apresentam, 
respectivamente, as especificações e a avaliação experimental  do CRDT proposto; a Secção \ref{sec: relatedWork} compara com trabalho relacionado; por \'ultimo
a Secção \ref{sec: conclusion} resume os contributos do artigo e aponta alguns t\'opicos a explorar no futuro.
\section{Conjunto remove\&add-wins CRDT}
\label{sec: especificacao}

\paragraph{\textbf{Especificação:}}
Para criar um conjunto remove\&add-wins CRDT é necessário sinalizar quais as operações de remoção que ganham sobre as de adição.
Uma possibilidade é estender a interface do conjunto com uma operação adicional de remoção, \remwinsope{}.
Assim, o conjunto remove\&add-wins CRDT tem uma interface com três operações:
\begin{inparaenum}[(i)]
\item \remwinsope{}, para remover o elemento $e$ com máxima prioridade;
\item \addope{}, para adicionar o elemento $e$; e
\item \remope{}, para remover o elemento $e$.
\end{inparaenum}

Para um dado conjunto de operações $O$, e considerando a relação de 
\emph{happens-before}
\cite{Lamport78Time} estabelecida entre as operações, é possível definir de modo preciso o estado do conjunto como sendo: \\
\vspace{-1.0mm}
$$
\begin{array}{ll}
\{\, e \mid \exists\, \mathit{add}(e) & \wedge \ (\nexists\, \mathit{remove}(e) \mid \mathit{add}(e) \prec \mathit{remove}(e))  \\ 
& \wedge \  (\forall\, \mathit{removeWins}(e) \mid \mathit{removeWins}(e) \prec \mathit{add}(e))\, \}
\end{array}
$$

Um elemento $e$ pertence ao conjunto se existir uma operação de \addope{} para a qual:
\begin{inparaenum}[(i)]
\item não existe nenhum \remope{} que tenha acontecido depois do \addope{}; e
\item todos as operações de \remwinsope{} aconteceram antes do \addope{}.
\end{inparaenum}

\paragraph{\textbf{Desenho:}}
Um dos
CRDTs presente na literatura que implementa a política \addwins{} é o 
Observed-Remove Set (OR-Set),
introduzido por Shapiro et. al. \cite{shapiro2011comprehensive}.
Neste CRDT, quando um elemento é adicionado, um identificador único é criado e associado ao elemento.
No \rem{}, todos os identificadores únicos associados a esse elemento que são conhecidos na replica fonte são marcados como removidos.
Com esta abordagem, se acontecer um \add{} e \rem{} concorrentes para o mesmo elemento, como o \rem{} não viu o identificador único associado ao \add{}, esse identificador único não é marcado como removido e, por isso, o elemento continua no conjunto.

A intuição por detrás do nosso conjunto remove\&add-wins CRDT é similar, mas estendendo a associação de identificadores únicos tanto ao \remwinsop{} como ao \addop{}. Assim, de modo a que o \remwinsop{} ganhe sobre um \addop{} concorrente, gera-se um identificador \remwinsop{} único no \remwinsop{} e marcam-se todos estes identificadores únicos conhecidos para remoção no \addop{}. Para que o \addop{} ganhe sobre um \remop{} concorrente, gera-se um identificador \addop{} único no \addop{} e marcam-se os identificadores únicos conhecidos para remoção no \remop{}.

Seguindo a especificação apresentada anteriormente, um elemento estará no conjunto se existir um identificador \addop{} único que não tenha sido marcado para remoção por um \remop{} e se não existir nenhum identificador \remwinsop{} único que não tenha sido marcado por remoção por algum \addop{}.

No Algoritmo \ref{alg:stateoarset} apresentamos uma especificação state-based não otimizada 
que implementa a referida ideia.
O estado é composto por:
\begin{inparaenum}[(i)]
\item mapa \emph{R} que mant\'em, para cada elemento, os identificadores únicos criados por uma operação \remwinsop{};
\item mapa \emph{A} que mant\'em, para cada elemento, os identificadores únicos criados por uma operação \addop{};
\item conjunto \emph{T} que mant\'em os identificadores únicos marcados para remoção quer por uma operação de \addop{} ou \remop{}.
\end{inparaenum}

A operação \remwinsope{} gera um identificador único e adiciona-o ao mapa \emph{R}, o qual é usado para que o \remwinsop{} ganhe sobre um \addop{} concorrente.

A operação \addope{} gera um identificador único e adiciona-o ao mapa \emph{A}, o qual é usado para que o \addop{} ganhe sobre um \remop{} concorrente.
Para além disso, também adiciona todos identificadores únicos associados ao elemento $e$ do mapa \emph{R} ao conjunto de identificadores únicos removidos (\emph{T}). Isto permite cancelar os efeitos dos \remwinsop{} que aconteceram antes do \addop{}.

A operação \remope{} adiciona todos identificadores únicos associados ao elemento $e$ do mapa \emph{A} ao conjunto dos identificadores únicos removidos (\emph{T}). Isto permite cancelar os efeitos dos \addop{} que aconteceram antes do \remop{}.

Nesta versão não otimizada, a operação de \emph{merge} simplesmente junta o estado dos mapas \emph{R} e \emph{A} e do conjunto \emph{T} de ambas as réplicas.

A operação de \emph{lookup}(\emph{e}) considera que um elemento \emph{e} está no conjunto se:
\begin{inparaenum}[(i)]
    \item todos os identificadores únicos em $R[e]$ tiverem sido removidos (estão em $T$);
    \item existe pelo menos um identificador único em $A[e]$ que não foi removido (ou seja, não está em $T$).
\end{inparaenum}

\begin{algorithm}[ht]
\caption{Conjunto remove\&add-wins state-based}
\label{alg:stateoarset}
\begin{algorithmic}[1]
\Payload{map $R$, map $A$, set $T$\\
$\quad R$: \emph{removeWins} elements $\mapsto$ set of unique ids\\
$\quad A$: \emph{add} elements $\mapsto$ set of unique ids \\
$\quad T$: removed unique ids}
\EndPayload
\Query{lookup}{element \emph{e}}{boolean}{b}
\State \Let \emph{b} = ($R[e] \setminus T = \varnothing \land A[e] \setminus T \neq \varnothing$)
\Comment{\footnotesize{Todos  \emph{removeWins} foram sobrepostos por \emph{add} e existe pelo menos um identificador único em \emph{A} que  não foi removido.}}
\EndQuery
\Update{add}{element \emph{e}}
    \State \Let $u$ = \emph{unique}() \Comment{\emph{unique}: gera um identificador único}
    \State $A[e].\mathit{add}(u)$
    \State $T.\mathit{add}(R[e])$
\EndUpdate
\Update{remove}{element \emph{e}}
    \State \Pre \emph{lookup}($e$) \Comment{Apenas remove o elemento se existir}
    \State $T.\mathit{add}(A[e])$
\EndUpdate
\Update{removeWins}{element \emph{e}}
    \State \Let $u$ = \emph{unique}() \Comment{\emph{unique}: gera um identificador único}
    \State $R[e].\mathit{add}(u)$
\EndUpdate
\Merge{X}{Y}{Z}
    \State \Let {$\forall e \!\in\! \mathit{dom}(X.R \cup Y.R) \! : Z.R[e] \! = \! X.R[e] \cup Y.R[e]$}
    \State \Let {$\forall e \!\in\! \mathit{dom}(X.A \cup Y.A) \! : Z.A[e] \! = \! X.A[e] \cup Y.A[e]$}
    \State \Let {$Z.T = X.T \cup Y.T$}
\EndMerge
\end{algorithmic}
\end{algorithm}

O conjunto remove\&add-wins  
state-based 
comporta-se como um conjunto sequencial para operações sequenciais. 
Operações concorrentes em elementos diferentes comutam naturalmente.
Para operações concorrentes no mesmo elemento, o \remwinsop{} ganha sobre \addop{} concorrentes que, por sua vez, ganha sobre \remop{} concorrentes.
A convergência eventual é garantida desde que o grafo formado pela sincronização das réplicas seja conexo -- nenhuma informação é alguma vez apagada.

\begin{algorithm}[htb]
	\caption{Conjunto remove\&add-wins op-based}
	\label{alg:opoarset}
	\begin{algorithmic}[1]
		\Payload{map $R$, map $A$\\
			$\quad R$: \emph{removeWins} elements $\mapsto$ set of unique ids\\
			$\quad A$: \emph{add} elements $\mapsto$ set of unique ids}
		\EndPayload
		\Query{lookup}{element \emph{e}}{boolean}{b}
		\State \Let \emph{b} = ($R[e] = \varnothing \land A[e] \neq \varnothing$)
		\Comment{Todos os \emph{removeWins} foram sobrepostos por \emph{add} e existe pelo menos um identificador único em \emph{A}.}
		\EndQuery
		\Update{add}{element \emph{e}}
		\AtSourceMultipleReturn{element \emph{e}}{unique-id \emph{u}, set \emph{TR}}
		\State \Let $u$ = \emph{unique}() \Comment{\emph{unique}: gera um identificador único}
		\State \Let $TR$ = $R[e]$
		\EndAtSourceMultipleReturn
		\Downstream{element \emph{e}, unique-id \emph{u}, set \emph{TR}}
		\State \Pre $\forall u \in TR: \mathit{removeWins}(e, u)$ foi entregue
		\Comment{Entrega causal é suficiente}
		\State $R[e].\mathit{remove}(TR)$
		\If{$R[e] == \emptyset$} 
			\State $A[e].\mathit{add}(u)$
		\EndIf
		\EndDownstream
		\EndUpdate
		\Update{remove}{element \emph{e}}
		\AtSourceMultipleReturn{element \emph{e}}{element \emph{e}, set \emph{TR}}
		\State \Pre \emph{lookup}($e$) \Comment{Apenas remove o elemento se existir}
		\State \Let $TR$ = $A[e]$
		\EndAtSourceMultipleReturn
		\Downstream{element \emph{e}, set \emph{TR}}
		\State \Pre $\forall u \in TR: \mathit{add}(e, u)$ foi entregue
		\Comment{Entrega causal é suficiente}
		\State $A[e].\mathit{remove}(TR)$
		\EndDownstream
		\EndUpdate
		\Update{removeWins}{element \emph{e}}
		\AtSourceReturn{element \emph{e}}{unique}{u}
		\State \Let $u$ = \emph{unique}() \Comment{\emph{unique}: gera um identificador único}
		\EndAtSourceReturn
		\Downstream{element \emph{e}, unique \emph{u}}
		\State $R[e].\mathit{add}(u)$ 
		\State $A[e].\mathit{removeAll}()$ 
		\EndDownstream
		\EndUpdate
	\end{algorithmic}
\end{algorithm}

A especificação op-based do conjunto remove\&add-wins encontra-se no Algoritmo \ref{alg:opoarset}.
A especificação usa a mesma aproximação da versão state-based, mas 
explora o facto das operações serem entregues por ordem causal para remover
imediatamente dos mapas \emph{R} e \emph{A} os identificadores únicos, dado que
em qualquer réplica uma operação que remova um dado identificador único é executada 
sempre após ter sido executada a operação que o criou.


\subsection{Versão otimizada}

A versão apresentada do conjunto remove\&add-wins não está otimizada, visto que os identificadores únicos que são adicionados nunca são removidos do estado do CRDT.
Seguindo a abordagem proposta por Bieniusa et. al. \cite{Bieniusa12Optimized} apresenta-se no Algoritmo \ref{alg:optstateoarset} a versão otimizada do conjunto remove\&add-wins.

\begin{algorithm}[!ht]
\caption{Conjunto otimizado remove\&add-wins state-based}
\label{alg:optstateoarset}
\begin{algorithmic}[1]
\Payload{map $R$, map $A$, vect v\\
$\quad R$: element $\mapsto$ set of pairs (timestamp t, replica r) 
\Comment{\footnotesize{identificadores \emph{removeWins}}}\\
$\quad A$: element $\mapsto$ set of pairs (timestamp t, replica r)
 \Comment{\footnotesize{identificadores \emph{add}}}\\
$\quad v$: summary (vector) of received pairs, initially [0,\ldots,0]}
\EndPayload
\Query{lookup}{element \emph{e}}{boolean}{b}
\State \Let \emph{b} = $A[e] \neq \varnothing$
\EndQuery
\Update{add}{element \emph{e}}
    \State \Let $r$ = \emph{myID}() \Comment{r: réplica fonte}
    \State \Let $t$ = ++v[r]
    \State $R[e].clear()$
    \State $A[e].\mathit{add}((t,r))$
\EndUpdate
\Update{remove}{element \emph{e}}
    \State \Pre \emph{lookup}($e$) \Comment{Apenas remove o elemento se existir}
    \State $A[e].clear()$
\EndUpdate
\Update{removeWins}{element \emph{e}}
    \State \Pre \emph{lookup}($e$) \Comment{Apenas remove o elemento se existir}
    \State \Let $r$ = \emph{myID}() \Comment{r: réplica fonte}
    \State \Let $t$ = ++v[r]
    \State $R[e].\mathit{add}((t,r))$
    \State $A[e].clear()$
\EndUpdate
\Merge{X}{Y}{Z}
	\For{$e \!\in\! \mathit{dom}(X.R \cup Y.R)$}
	    \State \Let {$M_X = \{(t,r) \in X.R[e] : Y.v[r] < t\}$}
	    \State \Let {$M_Y = \{(t,r) \in Y.R[e] : X.v[r] < t\}$}
	    \State \Let {$Z.R[e] = (X.R \cap Y.R) \cup M_X \cup M_Y $}
	\EndFor
	\For{$e \!\in\! \mathit{dom}(X.A \cup Y.A)$}
		\If{$e \!\not \in\! \mathit{dom}(Z.R) \vee Z.R[e] = \emptyset$}
	    \State \Let {$M_X = \{(t,r) \in X.A[e] : Y.v[r] < t\}$}
	    \State \Let {$M_Y = \{(t,r) \in Y.A[e] : X.v[r] < t\}$}
	    \State \Let {$Z.A[e] = (X.A \cap Y.A) \cup M_X \cup M_Y $}
	    \EndIf
	\EndFor
    \State \Let {$Z.v = [max(X.v[0],Y.v[0]),\dots,max(X.v[n-1],Y.v[n-1])]$}
\EndMerge
\end{algorithmic}
\end{algorithm}

O objetivo por detrás da versão state-based otimizada é que, à medida que as operações vão sendo executadas, os identificadores únicos sejam removidos, como na versão op-based. Seria, por isso, desejável que, na remoção, fosse possível apagar do estado os identificadores únicos que estão a ser removidos. Assim, o CRDT seria mais eficiente em termos de espaço e, possivelmente, na execução do \emph{merge}.


O problema de remover os identificadores únicos imediatamente é que quando se efetua a operação de
\emph{merge}, caso uma réplica contenha um identificador único e outra não, não é possível saber
se a réplica que não tem o identificador porque ainda não foi criado ou porque já foi removido.
Para resolver este problema, cada réplica pode manter um vetor \emph{v} que tem um ``sumário'' dos identificadores únicos observados. Assim, no \emph{merge}, pode-se consultar esse vetor para perceber se um certo identificador já foi ou não removido. 
Uma solução eficiente para implementar esse vetor é, para cada réplica, ter um contador: nesse caso os identificadores únicos passariam a ser um par (identificador da réplica, contador). O valor de entrada em cada posição do vetor \emph{v} será o maior valor de contador observado para a réplica nessa posição.

Em termos de alterações ao algoritmo do conjunto remove\&add-wins, com a alteração referida já não é 
necessário manter o conjunto dos identificadores removidos. Como tal, tanto o \remop{} como o \addop{} apagam logo os identificadores que devem ser removidos do conjunto \emph{A} e \emph{R}, respetivamente. O \addop{} e \remwinsop{} geram identificadores conforme discutido acima, atualizando também o vetor \emph{v}. O \emph{merge} junta os estados, tendo em conta quais os identificadores que já foram apagados através da informação dada pelo vetor \emph{v} em ambos os estados.

\section{Avaliação experimental}
\label{sec:avaliacao}

Nesta secção é avaliado experimentalmente o conjunto remove\&add-wins através da comparação com o conjunto OR, ambos na versão state-based otimizada. 
Nomeadamente, pretende-se avaliar qual o custo extra de utilizar o conjunto remove\&add-wins em termos de:
\begin{inparaenum}[(i)]
    \item tempo de execução; 
    \item espaço ocupado.
\end{inparaenum}

\vspace{-3mm}
\subsection{Características dos testes}
\label{subsec:caracteristicas}
As experiências foram realizadas através da simulação de um sistema com 3 clientes, cada um com uma cópia local do CRDT a ser testado.
Todos os clientes são executados na mesma máquina, não sendo por isso considerado nem o tempo de transferência nem a latência da rede.

Em cada teste, cada cliente realiza 4 milhões de operações modificadoras do estado (\addop{}, \remop{} ou \remwinsop{}) na sua réplica local. Sendo que cada operação escolhe, aleatoriamente, um elemento de entre um alfabeto de 20000 elementos. 
Cada cliente propaga o estado a outro cliente a cada \emph{x} operações executadas. 

De modo a poder avaliar qual o impacto dos metadados e computação extra do conjunto remove\&add-wins em relação ao conjunto OR, variou-se os seguintes parâmetros entre testes:

\begin{enumerate}
    \item Percentagem de operações de adição e remoção: foram consideradas as percentagens de, respetivamente, 50\%-50\% e 90\%-10\% (no caso do conjunto remove\&add-wins, metade das remoções são \remop{} e a outra metade \remwinsop{}). Estas definições permitem avaliar a diferença de desempenho entre adições e remoções;
    \item Frequência do envio de estado para \emph{merge}: testou-se tanto enviar o estado a cada 200000 operações (20 \emph{merges} no total) como apenas o enviar depois de o teste estar completo. Esta diferença permite avaliar qual o impacto no desempenho que tem realizar a operação de \emph{merge}. Destaca-se que é expectável que a operação de \emph{merge} seja lenta, visto que tem complexidade O(n), sendo \emph{n} o número de elementos na união dos conjuntos.
\end{enumerate}

Cada teste foi executado 5 vezes, de modo a obter resultados mais precisos.

\subsection{Resultados}

Nas subsecções seguintes serão apresentados e discutidos os resultados mais relevantes, não sendo por isso incluídas todas as combinações dos parâmetros referidos na Secção \ref{subsec:caracteristicas}.

Para cada gráfico, é apresentado o resultado médio dos 3 clientes, ao longo de 5 execuções. 
Por exemplo, o tempo de execução apresentado é a média do tempo de execução de cada cliente em cada execução, ou seja, uma média de 15 valores.


\subsubsection{Frequência de sincronização}

\vspace{-10pt plus 3pt minus 3pt}

\begin{figure}%
    \centering
    \subfloat{{\includegraphics[width = 0.47\textwidth]{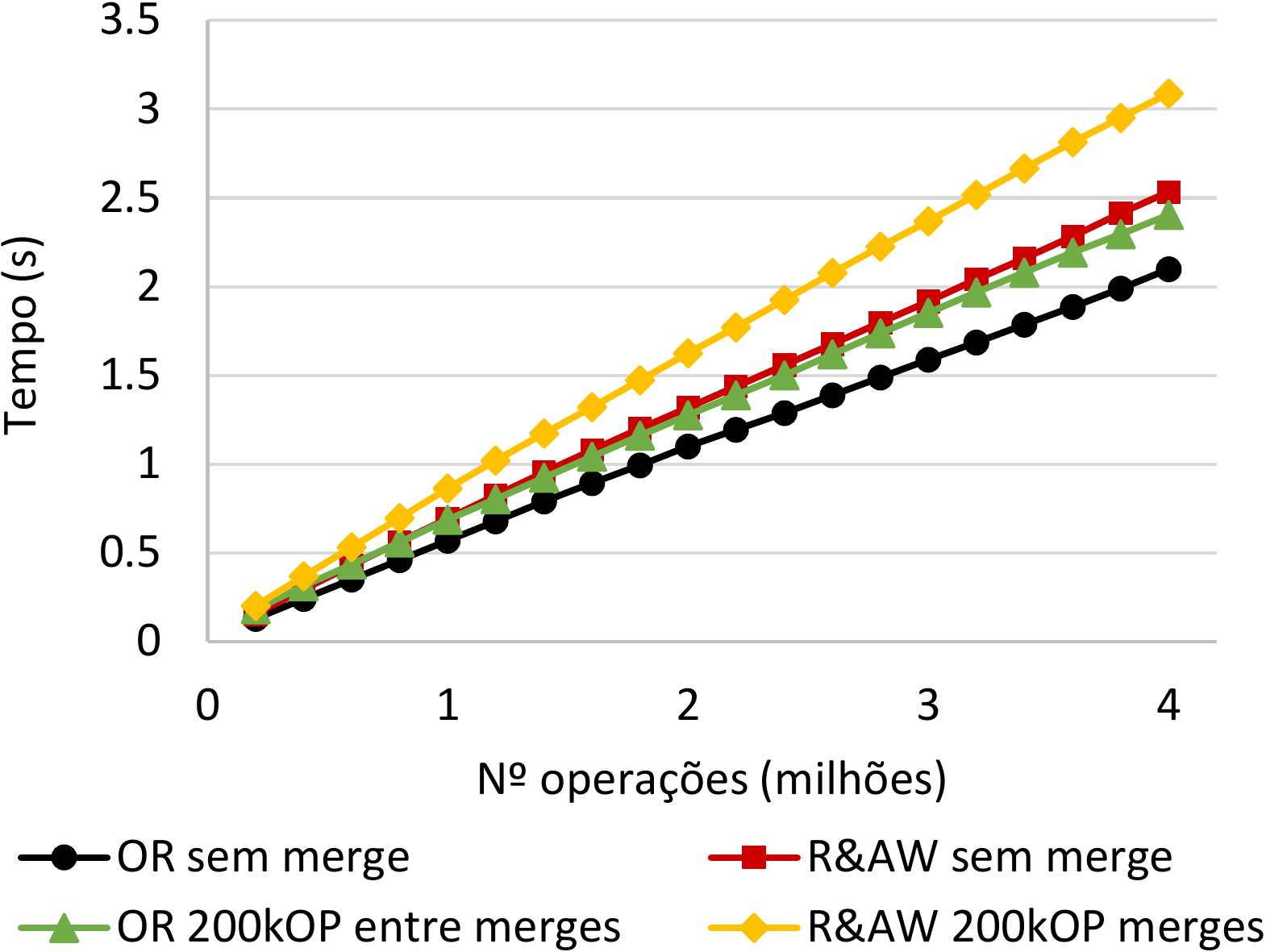}}}
    \qquad
    \subfloat{{\includegraphics[width = 0.47\textwidth]{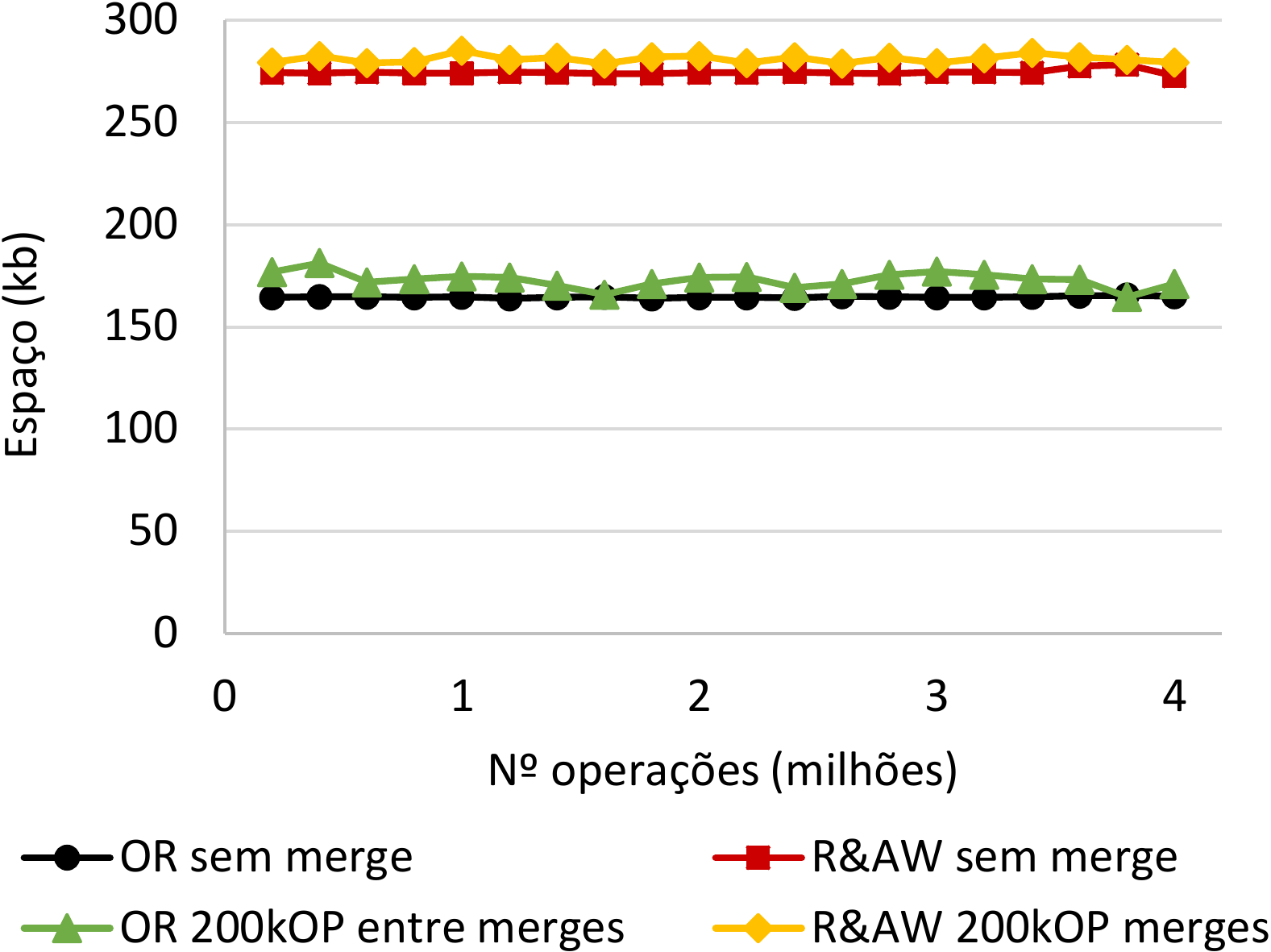}}}
    \caption{Tempo e tamanho dos metadados médios do conjunto OR e remove\&add-wins no teste com 50\% \addop{} e, respetivamente, 50\% \remop{} e 25\% \remop{} + 25\% \remwinsop{}.}
    \label{fig: graph_sync}
\end{figure}
\vspace{-3mm}
A Figura \ref{fig: graph_sync} apresenta o tempo de execução e o tamanho dos metadados médio de cada teste, para as percentagens de adição e remoção de 50\%-50\%. 

No gráfico do tempo, é possível verificar que existe uma diferença não desprezável entre o caso sem \emph{merge} e o caso com \emph{merge} a cada 200000 operações. Tal diferença deve-se ao custo do \emph{merge}, sendo esperado que se o \emph{merge} fosse mais frequente ou o número de elementos maior, a diferença também fosse maior.

Em relação ao espaço, este mantém-se relativamente constante - em média, cerca de metade dos elementos máximos (20000) estão no conjunto. Destaca-se que, conforme era esperado, o conjunto remove\&add-wins ocupa mais espaço em relação ao conjunto OR, devido à existência do mapa para guardar as remoções. Na prática, em média, para o remove\&add-wins cerca de 10000 elementos estão no mapa das adições e 5000 no das remoções (a percentagem de \remwinsop{} é de 25\%).





\subsubsection{Percentagem de adições e remoções}


\begin{figure}%
	\vspace{-15pt plus 5 pt minus 5 pt}
    \centering
    \subfloat{{\includegraphics[width = 0.47\textwidth]{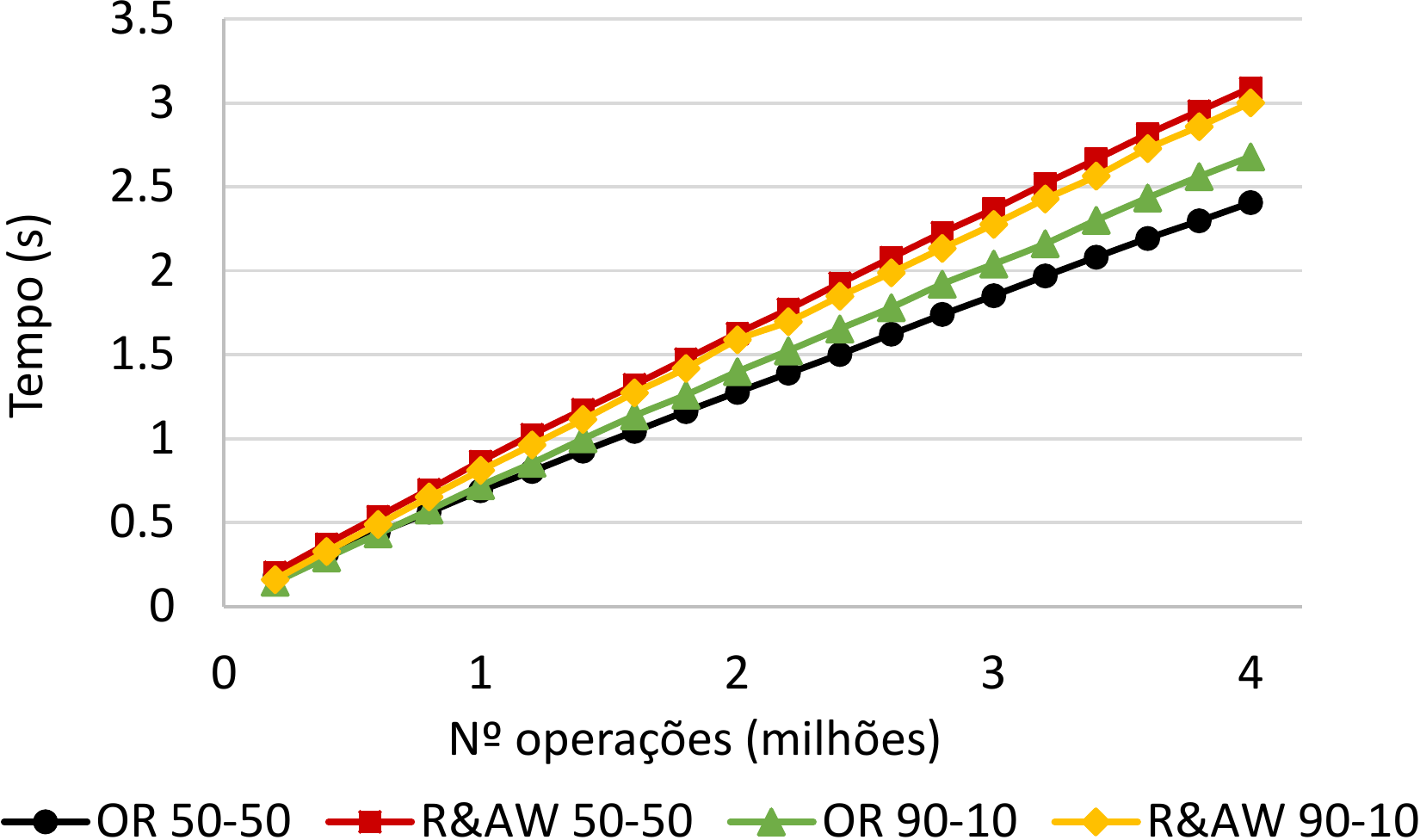}}}
    \qquad
    \subfloat{{\includegraphics[width = 0.47\textwidth]{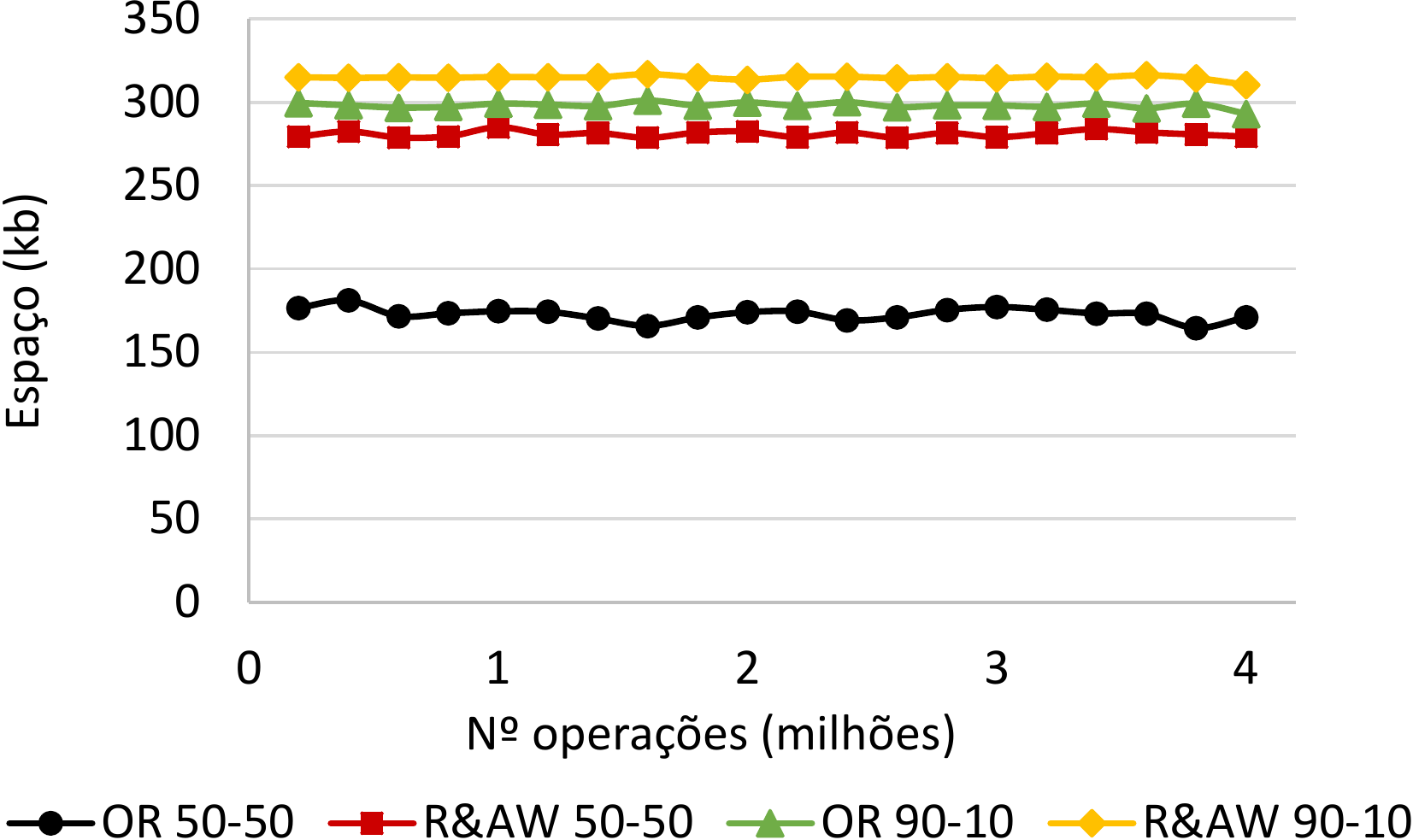}}}
    \caption{Tempo e tamanho dos metadados médios do conjunto OR e remove\&add-wins no teste com sincronização a cada 200000 operações, para as várias percentagens de operações.}
    \label{fig: graph_perc}
\end{figure}
\vspace{-2mm}
Os resultados relacionados com a diferença no tempo de execução e tamanho dos metadados para os casos de 50\% adições + 50\% remoções; 90\% adições + 10\% remoções encontram-se na Figura \ref{fig: graph_perc}. Em ambos os casos, os clientes sincronizam a cada 200000 operações.

Tanto no conjunto OR como no remove\&add-wins, é visível uma diferença no tempo de execução entre os casos de 50\%-50\% e 90\%-10\%. Essa diferença é, contudo, maior no caso do conjunto OR - tal facto deve-se a que as operações de \remwinsop{} e \addop{} executam computações que demoram, aproximadamente, o mesmo. Em ambos os conjuntos otimizados, a operação mais rápida é o \remop{}.

Em termos do espaço ocupado, no teste de 90\%-10\% existe uma tendência para estarem mais elementos no conjunto em relação ao de 50\%-50\%, o que leva a que mais espaço seja ocupado. No caso do conjunto remove\&add-wins, a diferença é menor devido aos elementos removidos por \remwinsop{} também serem guardados. Estes resultados permitem concluir que, quanto maior a percentagem de \addop{} (ou menor a de \remwinsop{}), menor a diferença entre os dois tipos de conjuntos no tamanho dos metadados - para 90\%-10\% a diferença é praticamente desprezável. 

\vspace{-2mm}
\subsubsection{Conjunto OR VS remove\&add-wins}

\begin{figure}%
    \centering
    \subfloat{{\includegraphics[width = 0.47\textwidth]{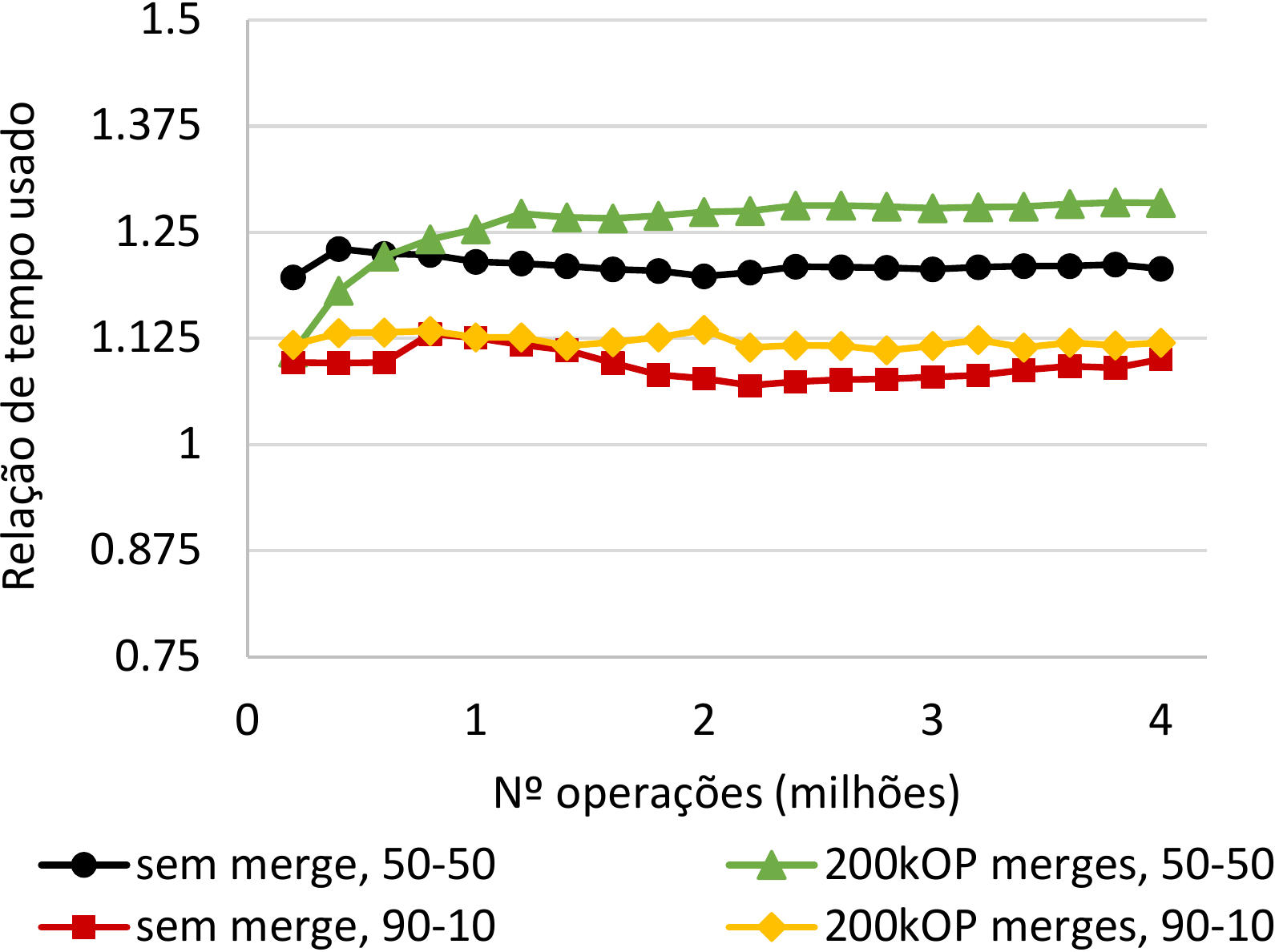}}}
    \qquad
    \subfloat{{\includegraphics[width = 0.47\textwidth]{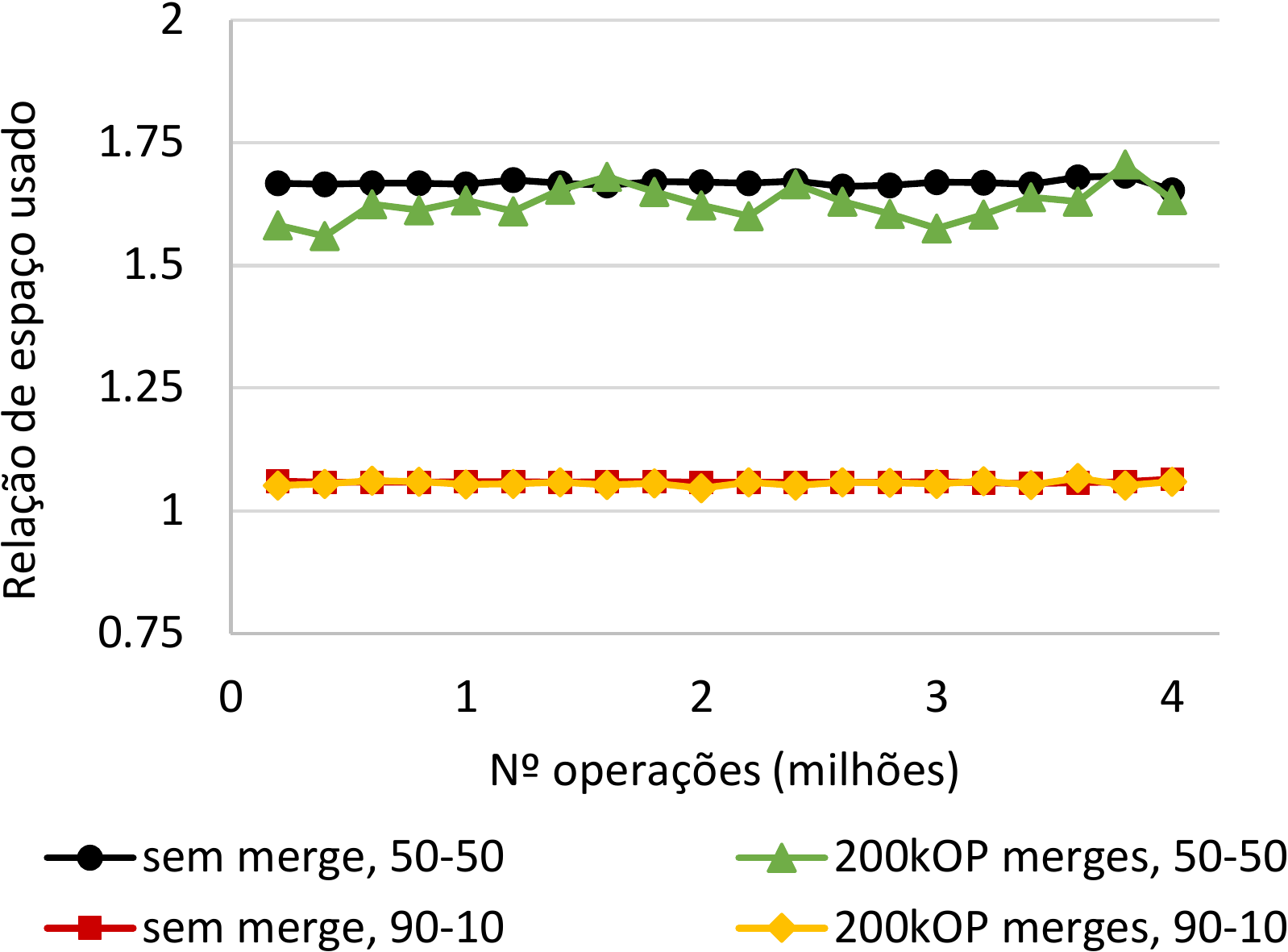}}}
    \caption{Relação de tempo e espaço usado pelos conjuntos. A relação é definida pela divisão do tempo/espaço do conjunto remove\&add-wins pelo conjunto OR.}
    \label{fig: graph_over}
\end{figure}

Na Figura \ref{fig: graph_over} encontram-se os resultados da divisão do tempo\textbackslash espaço do conjunto remove\&add-wins pelo OR.

Em termos do tempo de execução, o custo extra por utilizar o conjunto remove\&add-wins é relativamente baixo, sendo no pior caso cerca de 25\% extra. No caso do teste de 90\%-10\% (ou noutros casos com poucos \remwinsop{}), o custo extra é ainda menor (neste caso, 12.5\% extra). Contudo existirá sempre um custo extra devido ao passo extra que é necessário fazer no \addop{}. A diferença é maior no caso com sincronização, devido ao \emph{merge} ter que iterar mais um mapa.

Para o espaço utilizado, o custo extra é praticamente 
desprezável para o teste de 90\%-10\%, devido a existirem poucos elementos no mapa dos \remwinsop{}. No caso de 50\%-50\% a diferença é elevada, sendo no pior caso 70\% extra, devido aos elementos do \remwinsop{}. Espera-se que, mesmo com muitos \addop{}, se a percentagem de \remwinsop{} for baixa, o custo extra também seja baixo.

\vspace{-1mm}
\section{Trabalho relacionado}
\label{sec: relatedWork}
\vspace{-1mm}

Os princípios necessários para a especificação de tipos de dados replicados têm sido estudados intensivamente na literatura, tanto para ambientes com consistência forte \cite{wiesmann2000understanding} como para ambientes com consistência fraca 
\cite{burckhardt2013understanding, shapiro2011comprehensive}.

Entre as várias abordagens para a consistência fraca, destacam-se duas: OT (Operational Transformation) \cite{sun1998operational} e CRDTs (Conflict-Free Replicated Data Type) \cite{shapiro2011comprehensive}. O princípio do OT para lidar com conflitos de concorrência é de transformar os argumentos das operações remotas de modo a, depois de transformadas, as operações serem comutativas. Por outro lado, os CRDTs baseiam-se em tornar as operações comutativas desde o início.


Existem diversos CRDTs para representar diferentes tipos de dados, como por exemplo contadores, registos, conjuntos, mapas, listas ordenadas, top-K, etc \cite{shapiro2011comprehensive, preguica2009commutative, cabrita2017non}. 
Para o mesmo tipo de dados, pode existir várias especificações diferentes, variando no modo como são propagados os dados (state-based, op-based ou delta-based), \cite{shapiro2011comprehensive, almeida2014efficient}, na resolução de conflitos \cite{shapiro2011comprehensive, deftu2013scalable} e na eficiência \cite{deftu2013scalable, Bieniusa12Optimized}. Contudo, não é do nosso conhecimento que já tenha sido proposta alguma abordagem para ter num só CRDT várias políticas de resolução de conflitos.

O conjunto OR, no qual se baseia o nosso conjunto remove\&add-wins, foi proposto inicialmente por Shapiro. et.al. \cite{shapiro2011comprehensive}, enquanto que o princípio utilizado para a versão otimizada foi introduzido por Bieniusa et.al. \cite{Bieniusa12Optimized}.

\vspace{-1mm}
\section{Conclusão}
\label{sec: conclusion}
\vspace{-1mm}

Neste artigo discutiu-se o problema dos CRDTs apenas providenciarem uma única política para a resolução de conflitos.
Ilustrou-se o problema com base num cenário de exemplo no qual seria útil ter a opção de escolher entre mais que uma política de resolução de conflitos para o mesmo conjunto CRDT.

Propusemos uma solução para o referido cenário através da introdução de um conjunto CRDT que providência, através de duas operações de remoção diferentes, duas políticas para resolver conflitos: \addwins{} (\emph{sobre a remoção}) e \remwins{} (\emph{sobre a adição}).
Este CRDT dá ao programador a liberdade de selecionar qual a política que quer para cada cenário.
Usando uma aproximação semelhante, seria possível criar um conjunto CRDT que combina as duas políticas referidas através de duas operações de adição diferentes.

Realizou-se uma avaliação do desempenho do CRDT proposto em relação ao conjunto OR CRDT, concluindo-se que o custo extra é aceitável, sendo por vezes até desprezável se a percentagem de operações de \remwinsop{} for baixa.

De momento estamos a investigar uma aproximação mais genérica para este problema, na qual será possível especificar, na execução de uma operação, uma política definida pelo programador.




\vspace{-2mm}
\paragraph{\textbf{Agradecimentos:}}
Este trabalho foi parcialmente financiado pela FCT/MCTES através do projecto NOVA LINCS (UID/CEC/04516/2013) e a EU através do projecto LightKone (732505).


\vspace{-1mm}
\begin{thebibliography}{10}
\providecommand{\url}[1]{\texttt{#1}}
\providecommand{\urlprefix}{URL }
\providecommand{\doi}[1]{https://doi.org/#1}

\bibitem{almeida2014efficient}
Almeida, P.S., Shoker, A., Baquero, C.: Efficient state-based crdts by
  delta-mutation. In: International Conference on Networked Systems. pp.
  62--76. Springer (2015)

\bibitem{Bieniusa12Optimized}
Bieniusa, A., Zawirski, M., Pregui{\c c}a, N., Shapiro, M., Baquero, C.,
  Balegas, V., Duarte, S.: {An optimized conflict-free replicated set}. Rapport
  de recherche RR-8083, INRIA (Oct 2012),
  \url{http://hal.inria.fr/hal-00738680}

\bibitem{burckhardt2013understanding}
Burckhardt, S., Gotsman, A., Yang, H.: Understanding eventual consistency.
  Microsoft Research Technical Report MSR-TR-2013-39  (2013)

\bibitem{cabrita2017non}
Cabrita, G., Pregui{\c{c}}a, N.: Non-uniform replication. arXiv preprint
  arXiv:1711.07733  (2017)

\bibitem{deftu2013scalable}
Deftu, A., Griebsch, J.: A scalable conflict-free replicated set data type. In:
  Distributed Computing Systems (ICDCS), 2013 IEEE 33rd International
  Conference on. pp. 186--195. IEEE (2013)

\bibitem{gilbert2002brewer}
Gilbert, S., Lynch, N.: Brewer's conjecture and the feasibility of consistent,
  available, partition-tolerant web services. Acm Sigact News  \textbf{33}(2),
  51--59 (2002)

\bibitem{Lamport78Time}
Lamport, L.: Time, clocks, and the ordering of events in a distributed system.
  Communications of the ACM  \textbf{21}(7),  558--565 (1978)

\bibitem{preguica2009commutative}
Preguica, N., Marques, J.M., Shapiro, M., Letia, M.: A commutative replicated
  data type for cooperative editing. In: Distributed Computing Systems, 2009.
  ICDCS'09. 29th IEEE International Conference on. pp. 395--403. IEEE (2009)

\bibitem{shapiro2011comprehensive}
Shapiro, M., Pregui{\c{c}}a, N., Baquero, C., Zawirski, M.: A comprehensive
  study of convergent and commutative replicated data types. Ph.D. thesis,
  Inria--Centre Paris-Rocquencourt; INRIA (2011)

\bibitem{sun1998operational}
Sun, C., Ellis, C.: Operational transformation in real-time group editors:
  issues, algorithms, and achievements. In: Proceedings of the 1998 ACM
  conference on Computer supported cooperative work. pp. 59--68. ACM (1998)

\bibitem{wiesmann2000understanding}
Wiesmann, M., Pedone, F., Schiper, A., Kemme, B., Alonso, G.: Understanding
  replication in databases and distributed systems. In: Distributed Computing
  Systems, 2000. Proceedings. 20th International Conference on. pp. 464--474.
  IEEE (2000)

\end{thebibliography}
\end{document}